\documentclass[twocolumn,resetfootnote]{aastex7}
\usepackage{xcolor}
\usepackage{hyperref}
\usepackage{amssymb} 
\usepackage{multirow}
\usepackage{amsmath}

\begin{document}

\title{The Cocoon from a Massive Star's Death: VLA Radio Polarization Study of Possible Historical Supernova Remnant G7.7$-$3.7}

\author[orcid=0000-0002-9911-2509, gname=Tian-Xian, sname='Luo']{Tian-Xian Luo}
\affiliation{School of Astronomy and Space Science, Nanjing University, 163 Xianlin Avenue, Nanjing 210023, People's Republic of China}
\email{txluo@smail.nju.edu.cn}  

\author[orcid=0000-0002-5683-822X, gname=Zhou, sname='Ping']{Ping Zhou} 
\affiliation{School of Astronomy and Space Science, Nanjing University, 163 Xianlin Avenue, Nanjing 210023, People's Republic of China}
\affiliation{Key Laboratory of Modern Astronomy and Astrophysics, Nanjing University, Ministry of Education, People's Republic of China}
\email[show]{pingzhou@nju.edu.cn}


\author[orcid=0000-0002-5847-2612, gname=Ng, sname='C.-Y.']{C.-Y. Ng}
\affiliation{Department of Physics, The University of Hong Kong, Pokfulam Rd, Hong Kong, People's Republic of China}
\affiliation{Hong Kong Institute for Astronomy and Astrophysics, The University of Hong Kong, Pokfulam Road, Hong Kong, People's Republic of China}
\email{ncy@astro.physics.hku.hk}

\author[orcid=0000-0002-7299-2876, gname=Zhang, sname='Zhi-Yu']{Zhi-Yu Zhang} 
\affiliation{School of Astronomy and Space Science, Nanjing University, 163 Xianlin Avenue, Nanjing 210023, People's Republic of China}
\affiliation{Key Laboratory of Modern Astronomy and Astrophysics, Nanjing University, Ministry of Education, People's Republic of China}
\email{zzhang@nju.edu.cn}

\author[orcid=0000-0002-2096-6051, gname=Zhang, sname='Shumeng']{Shumeng Zhang}
\affiliation{Department of Physics, The University of Hong Kong, Pokfulam Rd, Hong Kong, People's Republic of China}
\affiliation{Hong Kong Institute for Astronomy and Astrophysics, The University of Hong Kong, Pokfulam Road, Hong Kong, People's Republic of China}
\email{shumeng_zhang@connect.hku.hk}

\author[orcid=0009-0004-6403-2981, gname=Hai-Chen, sname='Lin']{Hai-Chen Lin}
\affiliation{School of Astronomy and Space Science, Nanjing University, 163 Xianlin Avenue, Nanjing 210023, People's Republic of China}
\affiliation{Kuang Yaming Honors School, Nanjing University, 163 Xianlin Avenue, Nanjing 210023, People's Republic of China}
\email{221240028@smail.nju.edu.cn}

\begin{abstract}

G7.7$-3.7$ is a possible historical SNR, with the origin of its cocoon-like morphology and its supernova type remaining unclear. We performed L-band radio polarization observations of G7.7$-3.7$ using the Very Large Array in C and B-configurations. The high-resolution 1.4 GHz continuum image reveals a cocoon-like morphology with multiple shells and faint blowout structures.
The total flux density is $9.6\pm 0.5~{\rm Jy}$ and the spectral index map shows predominantly nonthermal emission, with an integrated spectral index of $-0.38\pm 0.04$. Polarization images of G7.7$-$3.7 show
high linear polarization fraction (30\%–40\%) in the northwestern filaments and moderate polarization
(10\%–20\%) in the northeast and south. The magnetic fields aligned with the filamentary structures, consistent with shock compression.
Large rotation measure (RM) variations across the SNR likely originate from magnetized massive progenitor winds. We suggest that the cocoon-like morphology results from the interaction between the SNR and pre-existing circumstellar shells, demonstrating that the radio polarization provides useful constraints on the environments and even the progenitor mass-loss.

\end{abstract}

\keywords{\uat{Supernova remnants}{1667} --- \uat{Radio continuum emission}{1340} --- \uat{Polarimetry}{1278} --- \uat{Stellar winds}{1636}}


\section{Introduction}

Supernova remnants (SNRs) are formed by interaction of supernova (SN) materials with the interstellar/circumstellar medium (ISM/CSM). They are key laboratories for studying plasma shock physics \citep{Bell2004, Ghavamian2007, Caprioli2014, Marcowith2016}, late stages of stellar evolution \citep{Dwarkadas2007, Dwarkadas2011, Patnaude2017, Patnaude2017b}, and the ambient ISM \citep{McKee1977, Reach2005, Sano2015, Sano2021}.

Radio observation is an essential window for studying SNRs, as most of them are visible in the radio band during their lifetimes. Many Galactic SNRs were first identified by radio observations and about 90$\%$ of the known Galactic SNRs were detected in the radio band \citep{Green2025}. Radio emission of SNRs comes from the synchrotron radiation emitted by the relativistic particles, which are accelerated by the shocks of SNRs and spiral around magnetic field lines. 

Synchrotron radiation is linearly polarized and its polarization direction (defined by the electric vector) indicates the magnetic field direction rotated by $90^{\circ}$. Therefore, through radio polarization observations, we can use the polarization fraction and polarization direction to probe the degree of ordering and orientation of the sky-projected magnetic field. This information can be used to diagnose physical processes near the shocks, such as magnetic field compression or turbulence, which modify the degree of ordering and orientation of the field \citep{Dubner2015,Slane2024a}.
Most SNRs show magnetic fields aligned with the shell. This magnetic field orientation is naturally explained as the shock compression of the ambient magnetic fields \citep{Dubner2015}.
However, radial magnetic fields have been found in a few youngest SNRs (Cas~A, Kepler's SNR, Tycho's SNR, SN~1006, RCW~89), implying turbulent magnetic amplification induced by cosmic rays or magnetohydrodynamic instabilities \citep{Reynoso2013a, Slane2024a, Zhang2025c}

The polarization direction can be rotated while the polarized radiation is traveling through magnetized plasma. This effect is called Faraday rotation and can be corrected by multi-frequency observations in radio band. In addition, the Faraday rotation measure (RM) can provide information of the line-of-sight magnetic field \citep{Brentjens2005, Ideguchi2022}. 
Notable RM variations have been observed across a few SNRs, such as G296.5+10.0 \citep{Harvey-Smith2010} and G1.9+0.3 \citep{Luken2020}. In both cases, these variations were proposed to originate from the CSM, which was shaped by the magnetized stellar wind of their progenitors. Therefore, mapping and analyzing RM in SNRs has the potential to study the wind materials of their progenitors prior to the SN events.



Historical SNRs are remnants of SNe that were seen with naked eyes and recorded in ancient texts. Since these SNRs are relatively bright in multi-wavelengths and connected with recorded SN events, the study of these remnants has much broadened our understanding of physics related to SNRs and SNe.

G7.7$-$3.7 is an intriguing SNR due to its peculiar cocoon-like morphology and the potential to be the remnant of a historical SN event \citep{Zhou2018b}. It was first identified as an SNR by \citet{Milne1974} based on its nonthermal spectrum. Using Parkes 2.7 GHz and 5 GHz polarization observations, \citet{Dickel1976} produced the first magnetic field and RM of G7.7$-$3.7. Subsequently, \citet{Milne1986} presented the Molonglo Observatory synthesis telescope (MOST) 843 MHz and Parkes 8.4 GHz observations of G7.7$-3.7$, revealing strong polarization at 8.4 GHz. \citet{Zhou2018b} proposed that G7.7$-3.7$ may be associated with the historical SN event SN 386 using \textit{XMM-Newton} observations, given its location within the ancient Chinese asterism Nan-Dou \citep{Stephenson2002}. This association implies that the SN that produced G7.7$-3.7$ was a low-luminous Type II SN. The first optical study of the remnant, conducted by \citet{Domcek2023b}, detected ${\rm H\,\alpha}+$[\ion{N}{2}], [\ion{O}{3}] and marginally [\ion{S}{2}] in its southern region, indicating an interaction between the SNR and a dense circumstellar shell. More recently, \citet{Cotton2024} presented the first high-resolution radio image of G7.7$-3.7$ using MeerKAT L-band observations, along with some polarization properties. However, these data suffer from significant flux loss of large-scale structures and cannot derive the polarization fraction. In addition, a more detailed discussion of the polarization properties of this remnant is needed 

Here, we present polarization observations of G7.7$-3.7$ in the L-band using the VLA. The details of the observations and the data reduction procedures are described in Section \ref{sec:data}. The observation results are presented in Section \ref{sec:results}. In Section \ref{sec:discussion}, we discuss the energetics, ambient environment, and RM variation of G7.7$-3.7$. Finally, a summary of our conclusions is provided in Section \ref{sec:conclusion}.

\section{Observation and Data Reduction} \label{sec:data}
\subsection{Observations}
We performed VLA observations of G7.7$-3.7$ in the L-band (at a central frequency of 1.49 GHz) with the C-configuration on February 3, 2020 and its B-configuration from June 24 to July 4, 2020 (PI: P.Zhou). The frequency coverage is 1008--1968 MHz and comprises 16 spectral windows (spws). Each spw comprises 64 channels with a channel width of 1 MHz. The observation was carried out using the full polarization mode with an integration time of 3 s. The summary of the observations is shown in Table \ref{table1}.

\begin{deluxetable*}{cccccccc}
\setlength{\tabcolsep}{2pt}
\tablewidth{0pt}
\tablecaption{Summary of VLA Observations of G7.7$-$3.7 \label{table1}}
\tablehead{
\colhead{Project} & 
\colhead{Date} & 
\colhead{\begin{tabular}{c} Array \\ Configuration \end{tabular}} & 
\colhead{\begin{tabular}{c} Central Frequency \\ (GHz) \end{tabular}} & 
\colhead{\begin{tabular}{c} Band Width \\ (MHz) \end{tabular}} & 
\colhead{\begin{tabular}{c} Total Time \\ (hr) \end{tabular}} &
\colhead{\begin{tabular}{c} Time on Target \\ (hr) \end{tabular}} &
\colhead{\begin{tabular}{c} Parallactic angle \tablenotemark{{\rm \dag}}  \\ ($^{\circ}$) \end{tabular}}
}
\startdata
20A-071 & 2020-07-04 & B & 1.49 & 960 & 2.2 & 1.5 & 19.3--40.1\\
20A-071 & 2020-07-02 & B & 1.49 & 960 & 2.2 & 1.5 & 8.6--32.7\\
20A-071 & 2020-06-24 & B & 1.49 & 960 & 2.2 & 1.5 & 10.6--34.1\\
20A-071 & 2020-02-03 & C & 1.49 & 960 & 2 & 1.4 & 323.4--344.0\\
\enddata

\tablenotetext{{\rm \dag}}{Parallactic angle range of complex gain calibrators J1833$-$2103 and J1811$-$2055.}
\end{deluxetable*}

For each observation, three calibrators were observed. 3C286 was used as flux, bandpass, and polarization angle calibrator. J1407+2827 as zero-polarization source was used as polarization leakage calibrator. J1833$-$2103 and J1811$-$2055 were used as complex gain calibrators and additional polarization leakage calibrators for C and B-configuration, respectively. 3C286 and J1407+2827 were observed once at the beginning of each observation. 
J1833$-$2103 and J1811$-$2055 were selected to be near G7.7$-$3.7 and were observed about every 15 minutes during observation of G7.7$-$3.7 for 7 scans and 8 scans, respectively.

\subsection{Calibration} 

Data reduction was carried out using the Common Astronomical Software Application (CASA) package \citep{Team2022}\footnote{\url{https://casa.nrao.edu/}}. We first applied Hanning smooth to continuum data to remove Gibbs ringing. Next, we performed a series of prior calibrations,  including corrections for antenna positions, gain-elevation curves, opacities, and Ionospheric Total Electron Content. Using the available model of the calibrator 3C286 \citep{Perley2017}, we derived and applied delay, flux density, and bandpass calibrations. J1833$-$2103 and J1811$-$2055 were used to derive and apply complex gain calibrations. To minimize the impact of radio frequency interference (RFI) and corrupted data, we used automated flagging algorithms (\texttt{tfcrop} and \texttt{rflag}) within the CASA task \texttt{flagdata}, supplemented by manual flagging when necessary during the calibration process. 

For polarization calibration, we started with the data processed through the continuum calibration steps described above, but without parallactic angle correction\footnote{\url{https://casaguides.nrao.edu/index.php/CASA_Guides:Polarization_Calibration_based_on_CASA_pipeline_standard_reduction:_The_radio_galaxy_3C75-CASA6.5.4}}. We set the 3C286 model with its known polarization properties\footnote{\url{https://science.nrao.edu/facilities/vla/docs/manuals/obsguide/modes/flux-density-scale-polarization-leakage-polarization-angle-tables}}. We then derived and applied the cross-hand delays using 3C286. 
Polarization leakage terms can be determined in two ways: either through a single observation of the unpolarized calibrator J1407+2827 or by using the unknown polarized calibrators J1833$-$2103 and J1811$-$2055, which also serve as complex gain calibrators in several scans. We solved the leakage terms in both two ways and yielded comparable solutions. We adopted the solutions from J1407+2827, except in the cases where its solutions were significantly affected by RFI, in which case the solutions derived from J1833$-$2103 and J1811$-$2055 were used.
Subsequently, the polarization angle was calibrated using 3C286, followed by the application of parallactic angle correction. Each observation was calibrated separately and finally combined using the CASA task \texttt{concat}. 

\subsection{Imaging}

After combing the C and B-configurations, the data were deconvolved interactively using the CASA task \texttt{tclean} to produce the final images. We adopted Briggs weighting with a robustness parameter of \texttt{robust=0.0} to balance sensitivity and resolution. We first produced a Stokes $I$ wideband image using multi-scale, multi-frequency synthesis, and applying a Gaussian taper in the uv-space with a half width at half maximum (HWHM) of 16 klambda (total uv-coverage: 0.1--22 klambda).
The reference frequency of the image was set to 1.4 GHz for convenience of flux density comparison. Then, the task \texttt{widebandpbcor} was applied to correct for the primary beam effects. The synthesis beam full width at half maximum (FWHM) of the final image is $7.6''\times6.2''$ with a position angle (PA) of $17^\circ$ and the root mean square (rms) noise of the image is ${\sim 10~\rm \mu Jy/beam}$.

To derive the RM distribution, subband images are also needed. We produced Stokes $I$, $Q$, $U$, and $V$ images for each spw using multi-scale synthesis with \texttt{nterms=1}. For subsequent processing and to boost signal to noise ratio (S/N), all subband images share the same pixel grid and were convolved to a common beam size of $38''$ FWHM. We applied the \texttt{impbcor} task to correct for primary beam effects and masked all the images to match the smallest primary beam size. The detailed frequency sets and rms noises of these subband images are listed in Table \ref{spw}. 

\begin{deluxetable}{cccl}
\tablewidth{0pt}
\tablecaption{Central frequencies and rms noises of subband images \label{spw}}
\tablehead{
\colhead{spw} & 
\colhead{\begin{tabular}{c} Frequency \\ (MHz) \end{tabular}} &
\colhead{\begin{tabular}{c} rms \\ (${\rm mJy~beam^{-1}}$) \end{tabular}} &
\colhead{Comment}
}
\startdata
0 & 1040 & 0.66 & \\
1 & 1104 & 0.71 & \\
2 & 1168 & -- & Blanked \tablenotemark{{\rm \dag}}\\
3 & 1232 & -- & Blanked\\
4 & 1296 & 0.65 & \\
5 & 1360 & 0.62 & \\
6 & 1424 & 0.58 & \\
\hline
7 & 1488 & \multirow{2}*{0.71} & \multirow{2}*{Identical \tablenotemark{{\rm \ddag}}}\\
8 & 1488 & ~ & ~\\
\hline
9 & 1552 & -- & Blanked\\
10 & 1616 & 2.1 & \\
11 & 1680 & 1.4 & \\
12 & 1744 & 1.4 & \\
13 & 1808 & 1.4 & \\
14 & 1872 & 1.4 & \\
15 & 1936 & 2.0 & \\
\enddata

\tablenotetext{{\rm \dag}}{These spws are blanked by RFI.}
\tablenotetext{{\rm \ddag}}{These two spws are identical since they are located at the edges of the two basebands, respectively.}

\end{deluxetable}


\subsection{RM synthesis}
We performed RM synthesis \citep{Brentjens2005} to derive polarization properties. 
This technique identifies, for each pixel, the RM value that maximizes the weighted average of the RM-corrected linear polarized intensity $P=\sqrt{Q^2+U^2}$. We took the highest peak $P$ as the linear polarized intensity, the corresponding RM value as the RM, and the corresponding RM-corrected $Q$ and $U$ to calculate the electric-vector polarization angle (EVPA) at zero wavelength $\chi_{0}=0.5{\rm arctan2}(U, Q)$. We also extracted the Faraday depth spectra for G7.7$-$3.7 and noticed that, in addition to the highest peak, there are some regions that have multiple local peaks in the Faraday depth spectra. A detailed discussion is provided in the Appendix \ref{sec:appendix}.

We implemented RM synthesis using RM-Tools \citep{Purcell2020}\footnote{\url{https://github.com/CIRADA-Tools/RM-Tools}}, searching over an RM range of $\pm 400~{\rm rad~m^{-2}}$. We adopted a optimal frequency weighting scheme based on the inverse variance ($1/\sigma^2$) of each spw \citep{Heald2009}. The debiased linear polarized intensity is calculated following \citet{Wardle1974, George2012} using the relation:
\begin{equation}
P_{\rm debiased}= \sqrt{Q^{2}+U^{2}-2.3\sigma_{QU}}
\end{equation}
applied to pixels where the polarized intensity exceeds $5\sigma_{QU}$. Here, $\sigma_{QU}$ denotes the average rms noise in the RM-corrected Stokes $Q$ and $U$ maps.

The reference frequency for the RM-synthesis results is 1310 MHz, obtained as the weighted average of the squared wavelengths (see Eq. 32 in \citealt{Brentjens2005}). The linear polarization fraction is calculated as $FP={P_{\rm debiased}}/{I_{1310~{\rm MHz}}}$, where $I_{1310~{\rm MHz}}$ denotes the Stokes I value rescaled to the reference frequency of 1310 MHz by fitting subband images to a power law ($S\propto \nu^{\alpha}$).

\subsection{Other Data}
We used the GaLactic and Extragalactic All-sky Murchison Widefield Array (GLEAM) survey at 80–300 MHz \citep{Hurley-Walker2017, Hurley-Walker2019b, Duchesne2025} to calculate the integrated spectral index. We also used Wide-field Infrared Survey Explorer (WISE) 22 ${\rm \mu m}$ All-Sky Survey data \citep{Wright2010a} for comparison. This survey was carried out in 2010 with an angular resolution of $12''$. 

\section{Results} \label{sec:results}

\subsection{Radio Morphology and Flux Density}
\begin{figure*}[ht!]
\centering
\resizebox{0.7\textwidth}{!}
{\plotone{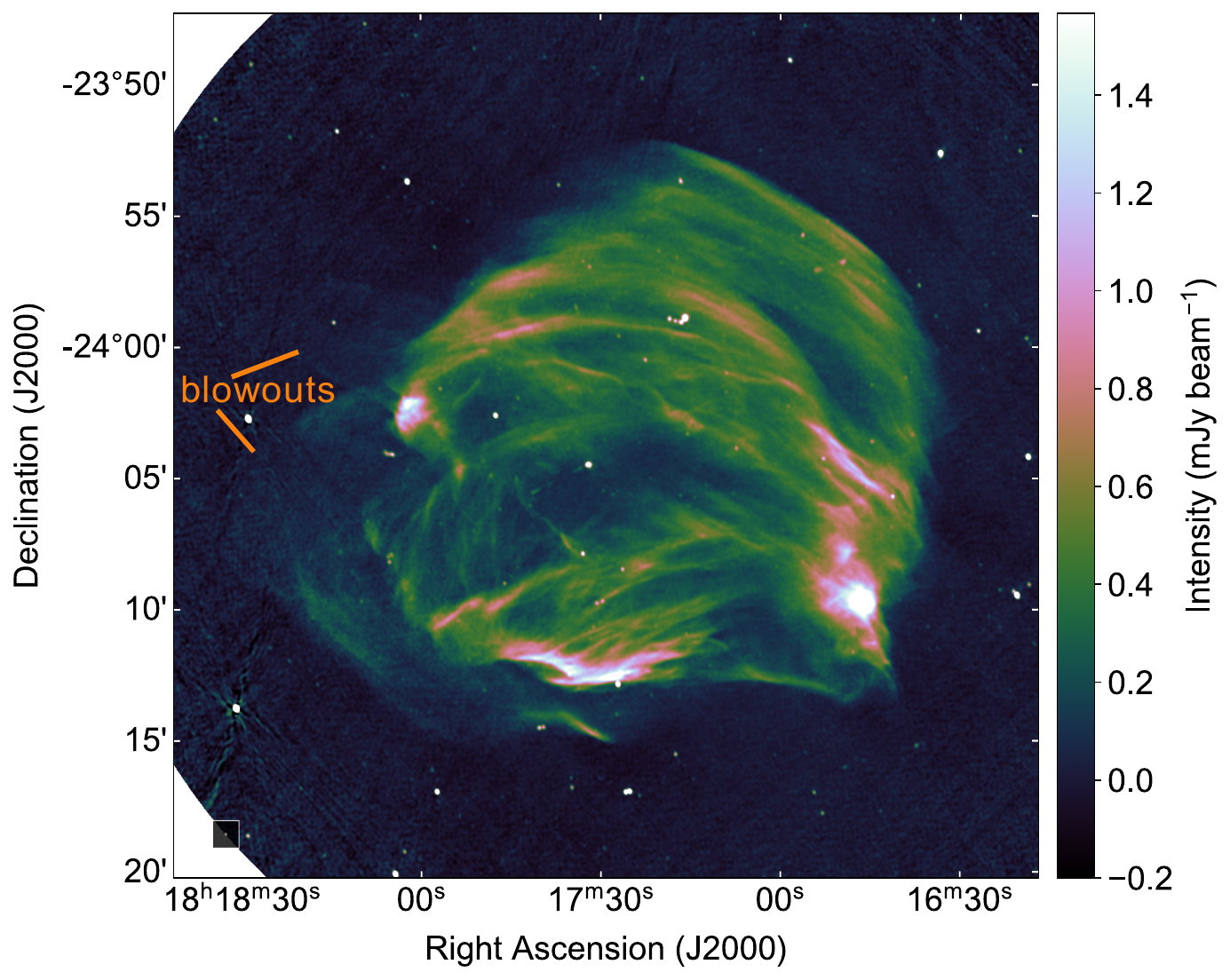}}
\caption{The wideband continuum intensity map of SNR G7.7$-$3.7 from B- and C-configurations of VLA at the reference frequency of 1.4 GHz with a bandwidth of 960 MHz. The gray ellipse in the left bottom corner denotes the synthesized beam ($7.6''\times6.2''$ FWHM, PA=$17^\circ$).
\label{fig:I}}
\end{figure*}

Figure \ref{fig:I} presents the wideband continuum image of SNR G7.7$-$3.7 at 1.4 GHz. The image, with a synthesized beam of $7.6'' \times 6.2''$ (PA = $17^\circ$), reveals the highly-structured morphology of the remnant. The morphology is characterized by two bright regions in the east and the southwest, a bright southern filament, multiple thin filaments in the northwest, and two faint blowouts in the eastern and southeastern regions, respectively. 

To calculate the flux densities $S$ of G7.7$-$3.7, we first used the Aegean \citep{Hancock2012, Hancock2018a} to detect point sources in the field of view and subtract their flux densities. Then we draw a polygon region to enclose G7.7$-$3.7 to calculate the total flux densities of the remnant region and another polygon region around it to calculate the average sky background. The flux density of G7.7$-$3.7 at 1.4~GHz is $9.6\pm0.5~{\rm Jy}$ derived using the total flux densities of the remnant region to subtract the average background flux densities. The total uncertainty is determined by $\Delta S=\sqrt{\sigma_{\rm sky}^{2}+\sigma_{\rm cal}^{2}}$, where the systematic uncertainty $\sigma_{\rm cal}$ is 5\% due to the calibration uncertainty\footnote{\url{https://science.nrao.edu/facilities/vla/docs/manuals/oss}}, and $\sigma_{\rm sky}$ is the uncertainty due to the flux fluctuation of the radio sky. $\sigma_{\rm sky}$ is derived as \citep{Klein1981, Cotton2024}:
\begin{equation}
\sigma_{\rm sky}= \sigma \sqrt{\frac{N^{2}_{\rm source}}{N_{\rm bg}}+N_{\rm source}}
\end{equation}
where $\sigma$ is the rms in Jy/pixel of the image, $N_{\rm source}$ and $N_{\rm bg}$ are the pixel numbers in  G7.7$-$3.7 and background regions.

The flux density we derived ($9.6\pm0.5~{\rm Jy}$) is significantly higher than the value of $5.2\pm 0.4~{\rm Jy}$ at 1335 MHz reported by \citet{Cotton2024} using MeerKAT observations, but agrees well with the $9.9\pm 0.1~{\rm Jy}$ expected at 1.4 GHz by \citet{Dubner1996} using VLA and the 30 m antenna of the Instituto Argentino de Radioastronomía (IAR). 
The largest angular scale (LAS) detectable in our observation is $16'$, smaller than the diameter of $22'$ for G7.7$-$3.7, thereby our radio observations may miss the very extended emission. Nevertheless, the radio emission of G7.7$-$3.7 is highly structured and the total flux of our image is consistent with that obtained by \citet{Dubner1996}, which is expected to be reliable since the data were made with a combination of an interferometer and a single dish telescope.
This consistency justifies that the VLA image at 1.4~GHz is not significantly influenced by the missing flux issue.

\subsection{Spectral Index}
We used the GLEAM survey centered at 88, 118, 155, 200, and 300 MHz, together with our measured flux density at 1.4 GHz, to derive the integrated spectral index of G7.7$-$3.7. The GLEAM survey was conducted using the Phase I configuration, with a shortest baseline of 7.72 m, providing sensitivity to angular scales up to $\sim 7.4^{\circ}$ at 300 MHz \citep{Duchesne2025}. Since the angular extent of G7.7$-$3.7 ($\sim 22'$) is much smaller than the maximum recoverable scales, we do not expect significant missing flux in the GLEAM images. 

The low-frequency images of GLEAM have non-uniform diffuse background emission, which can contaminate the flux density measurements and bias the spectral index determination. To mitigate this effect, we subtracted large-scale background emission using the \texttt{FINDBACK} command from the Starlink software package \citep{Currie2014, Berry2022} prior to calculating the spectral index. Because of the limited angular resolution of the GLEAM images, individual point sources could not be reliably subtracted. However, we estimate the total contribution from the bright point sources within the G7.7$-$3.7 region using the VLA 1.4 GHz image and the in-band spectral index map generated through multi-frequency \texttt{tclean}. The total contribution of these point sources is less than 0.4 Jy in the GLEAM frequencies, which is negligible compared to the uncertainties of flux densities. The calibration uncertainty $\sigma_{\rm cal}$ of GLEAM is 13\% for 88--200 MHz \citep{Hurley-Walker2019b} and 12\% for 300 MHz \citep{Duchesne2025}. 

\begin{figure}[ht!]
\centering
\resizebox{0.4\textwidth}{!}
{\plotone{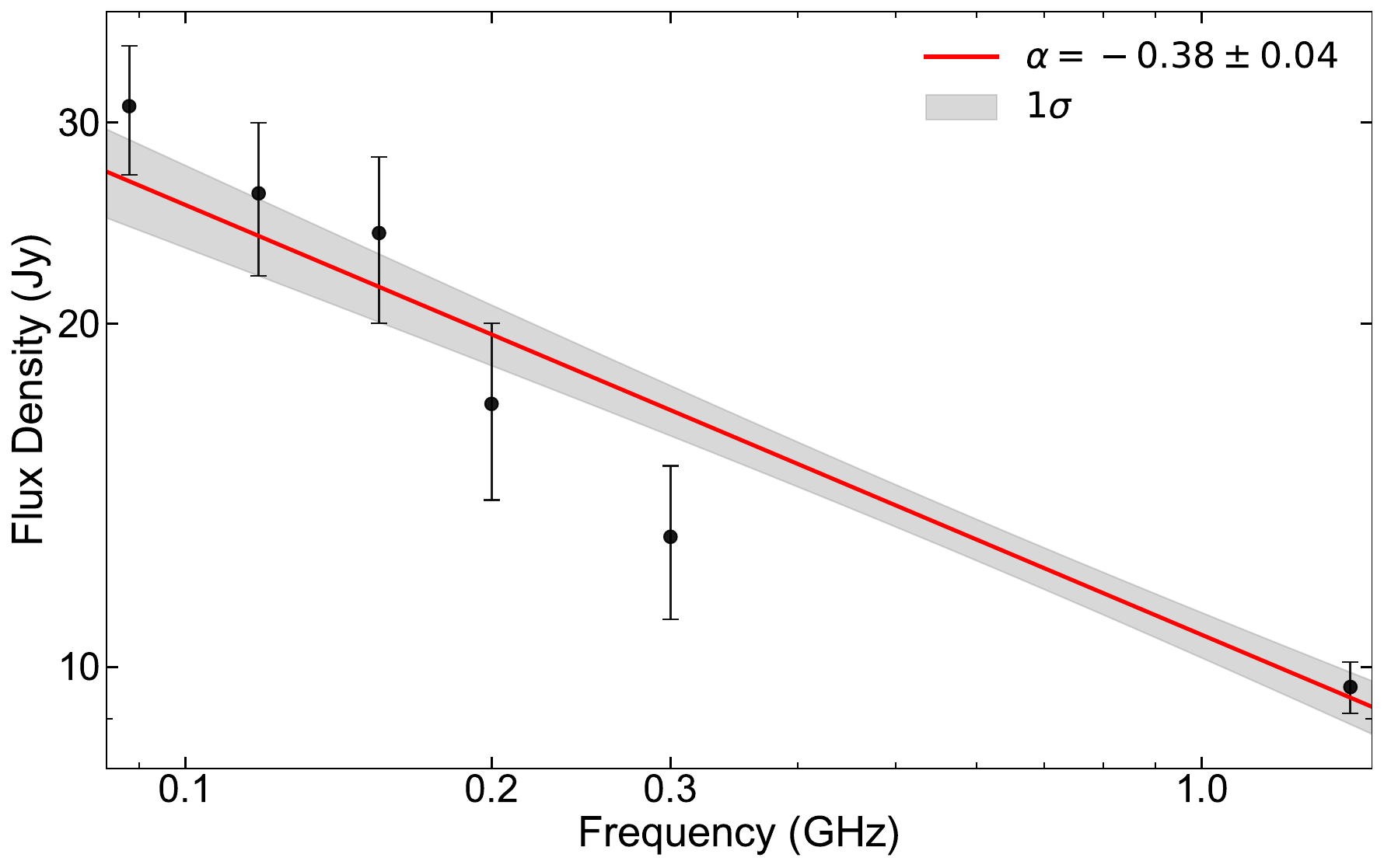}}
\caption{The flux intensity spectrum of G7.7$-$3.7. The black dots denote the flux density measured in 88, 118, 154, 200, and 300 MHz using GLEAM data and 1.4 GHz using our VLA observation, with 1-$\sigma$ error bars. The red line is the best-fit line with 1-$\sigma$ gray shadow.
\label{fig:spectrum}}
\end{figure}

Figure \ref{fig:spectrum} shows the integrated flux intensity spectrum of G7.7$-$3.7 derived from the GLEAM data and our VLA observation. A power-law fit yields an integrated spectral index of $\alpha = -0.38 \pm 0.04$, consistent with previous single-dish measurements \citep{Milne1974, Milne1986}.

\begin{figure}[ht!]
\centering
\resizebox{0.46\textwidth}{!}
{\plotone{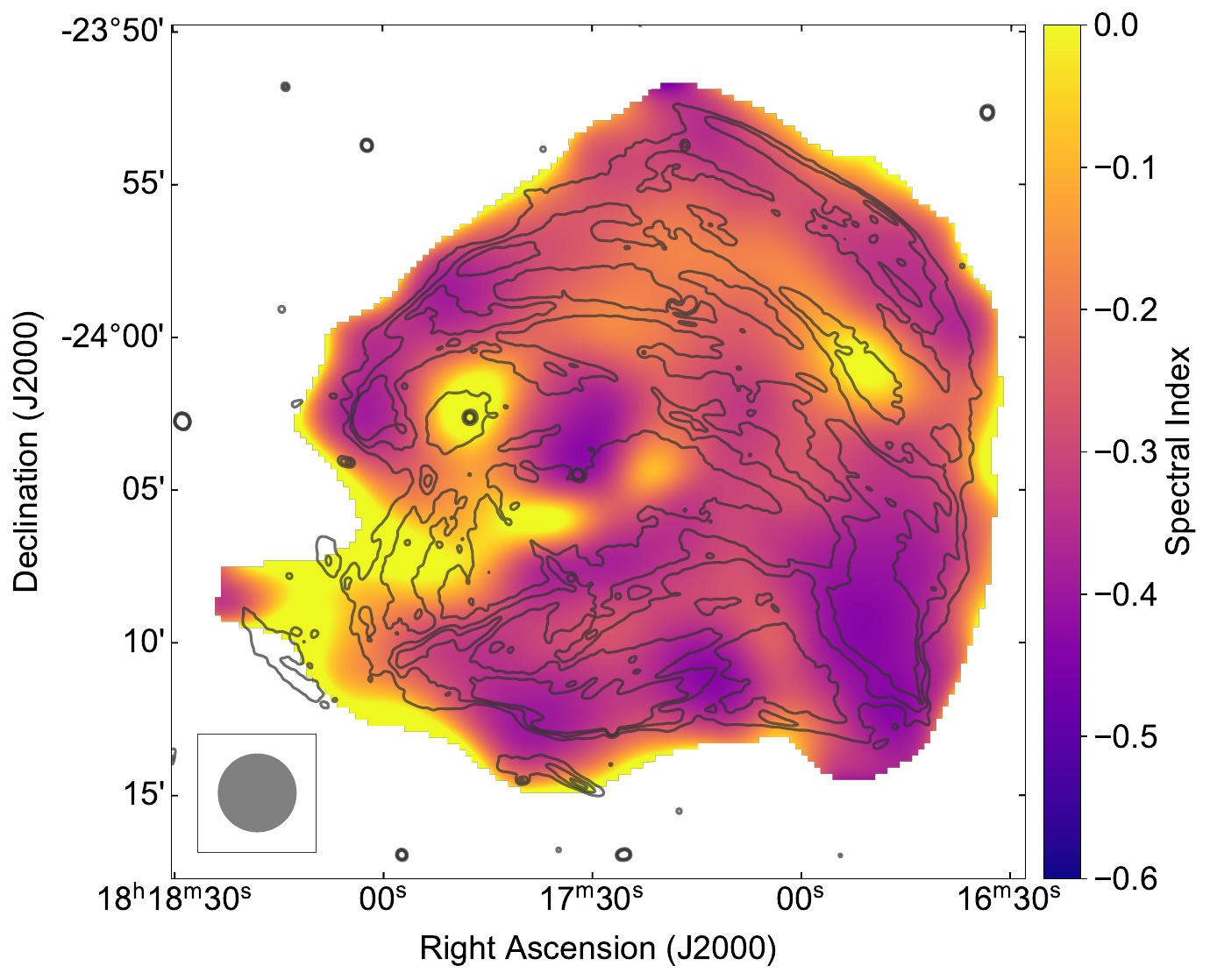}}
\caption{The spectral index map of G7.7$-$3.7 from 200 MHz to 1.4 GHz overlaid with gray contours of 1.4 GHz total intensity levels at 0.2 mJy/beam, 0.4 mJy/beam, and 0.6 mJy/beam. Pixels in which the flux density is less than $6\sigma$ are masked. The gray ellipse in the left bottom corner denote the smoothed beam ($155''\times155''$ FWHM).
\label{fig:spx}}
\end{figure}

To obtain a spatially resolved spectral index map with adequate angular resolution, we used only the 200 MHz, 300 MHz, and 1.4 GHz images. These images were convolved to a common angular resolution of $155''$ FWHM (the lowest resolution of these images) before fitting the spectral index pixel-by-pixel. The resulting spectral index map of G7.7$-$3.7 is shown in Figure~\ref{fig:spx}. The remnant exhibits a spectral index of $\sim -0.3$, indicating predominantly non-thermal emission.

\subsection{Polarization Properties}
\begin{figure*}[ht!]
\centering
\resizebox{0.99\textwidth}{!}
{\plotone{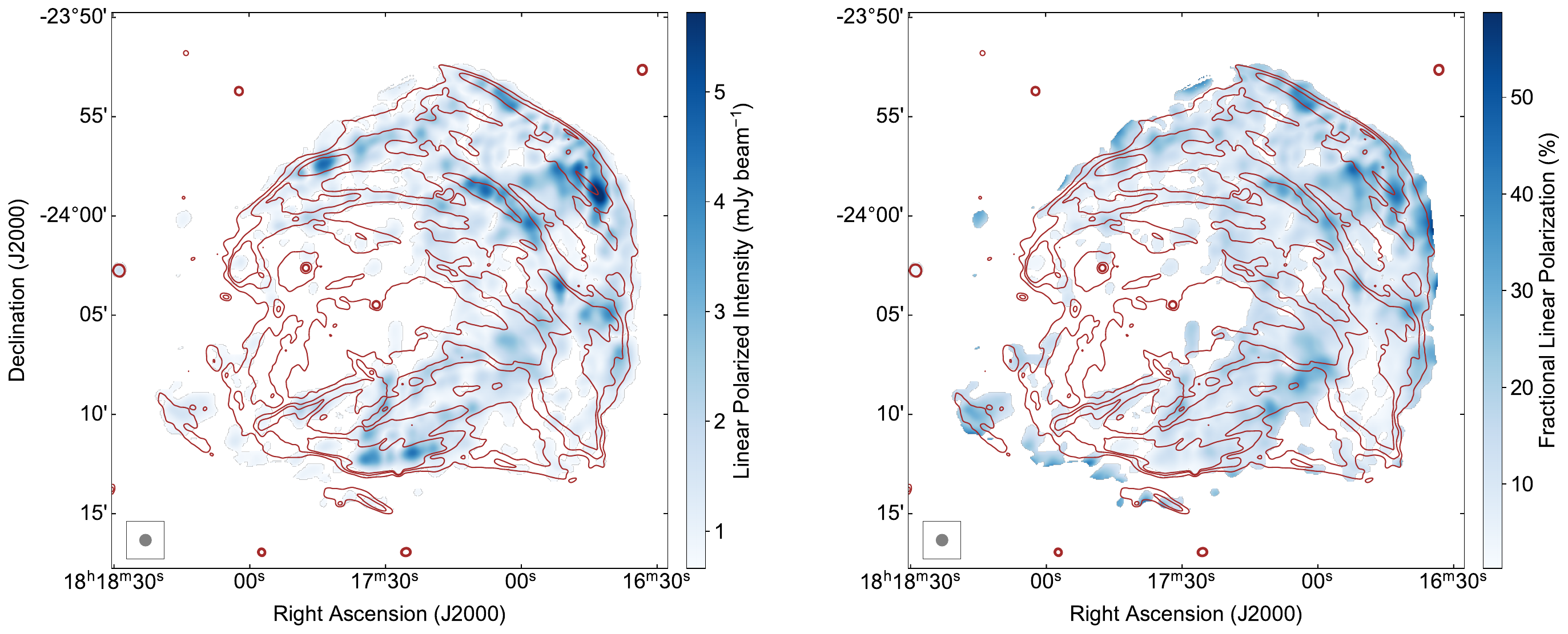}}
\caption{\textit{Left:} The linear polarized intensity of G7.7$-$3.7. \textit{Right:} The linear polarization fraction of G7.7$-$3.7. Pixels in which the linear polarization intensity are less than 2$\sigma$ or the subband Stokes I images are less than 6$\sigma$ are masked. The brown contours are total intensity levels at 0.2 mJy/beam, 0.4 mJy/beam, and 0.6 mJy/beam. The gray ellipse in the left bottom corner denote the synthesized beam ($38''$ FWHM).
\label{fig:PIPD}}
\end{figure*}

The linearly polarized intensity and the linear polarization fraction for G7.7$-$3.7 are presented in Figure \ref{fig:PIPD}. The northwestern filaments exhibit strong polarized intensity and a high linear polarization fraction, reaching approximately $30\%$--$40\%$, indicating that the magnetic fields are more ordered than in other regions. The northeastern filaments and the southern filaments also show a significant polarized intensity, but with a moderate linear polarization fraction of around $10\%$--$20\%$. No significant polarized emission was detected in the two bright regions located in the eastern and southwestern areas.

\begin{figure*}[ht!]
\centering
\resizebox{0.99\textwidth}{!}
{\plotone{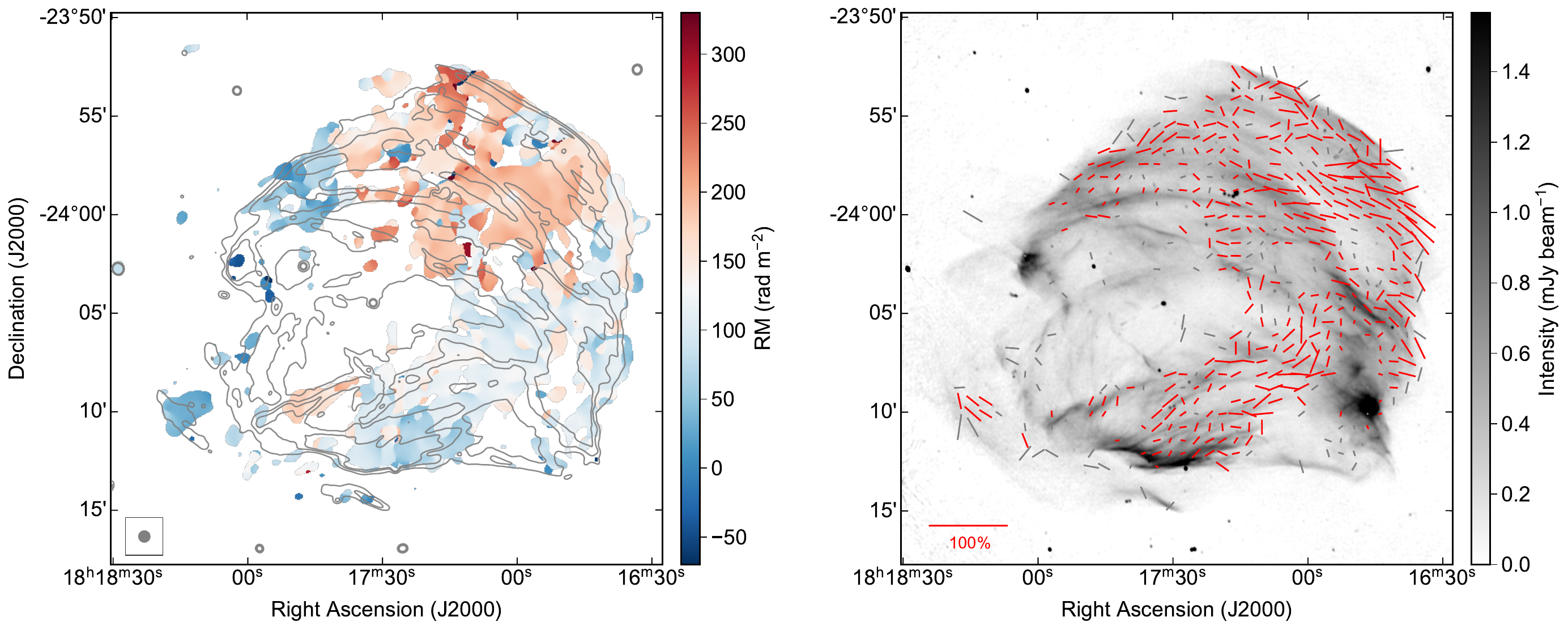}}
\caption{\textit{Left:} The RM map of G7.7$-$3.7 overlaid with gray contours of total intensity levels at 0.2 mJy/beam, 0.4 mJy/beam, and 0.6 mJy/beam. Pixels in which the linear polarization intensity are less than 3$\sigma$ and the error of RM are more than 10 ${\rm rad~m^{-2}}$ are masked. The gray ellipse in the left bottom corner denote the synthesized beam ($38''$ FWHM). \textit{Right:} Sky-projected magnetic field vectors overlaid on the L-band continuum intensity map of G7.7$-$3.7. The vector direction corresponds to the sky-projected magnetic field orientation, obtained by rotating the EVPA at zero wavelength by $90^\circ$. The vector length scales with the linear polarization fraction. A reference vector for 100\% polarization is shown in red in the bottom left corner. The same masking threshold as applied in the linear polarization fraction map (Figure \ref{fig:PIPD}) is used for these vectors. The red vectors indicate the linear polarization intensity are more than 3$\sigma$.
\label{fig:RMPA}}
\end{figure*}

The RM map and the sky-projected magnetic field vectors overlaid on the L-band continuum intensity map of G7.7$-$3.7 are shown in Figure \ref{fig:RMPA}. The RM across G7.7$-3.7$ is predominantly positive and exhibits a clear gradient. The northwest region shows a high RM value of $\sim 180~{\rm rad~m^{-2}}$, while the southwest region has a lower RM ranging from $70$ to $130~{\rm rad~m^{-2}}$. A small region in the northeast displays a low RM of $\sim 60~{\rm rad~m^{-2}}$, and a filament in the southeast reaches a high RM of roughly $170~{\rm rad~m^{-2}}$. The bright eastern region is characterized by a negative RM value of $\sim -39~{\rm rad~m^{-2}}$. 
The sky-projected magnetic field directions are mostly aligned with the filaments of the remnant in the northern, western, and southern regions. The RM and EVPA distributions agree with the MeerKAT results \citep{Cotton2024}.

~\\

The summary of our observation results, including the flux density, spectral index and polarization results, are shown in Table \ref{results}.

\begin{deluxetable}{cc}
\tablewidth{0pt}
\tablecaption{Summary of VLA observation results of G7.7$-$3.7 \label{results}}
\tablehead{
\colhead{Parameter} & 
\colhead{Value}
}
\startdata
Flux density & $9.6\pm0.5~{\rm Jy}$ (1.4 GHz)\\
integrated spectral index & $-0.38\pm0.04$ \\
linear polarization fraction range & $\sim 2$--$46\%$\\
Average linear polarization fraction & $\sim 14\%$\\
Rotation measure range & $\sim -35$--$211~{\rm  rad~m^{-2}}$\\
Average rotation measure & $\sim 130~{\rm rad~m^{-2}}$\\
\enddata
\end{deluxetable}

\section{Discussion} \label{sec:discussion}

\subsection{The ambient environment} \label{sec:en}
\begin{figure}[ht!]
\centering
\resizebox{0.45\textwidth}{!}
{\plotone{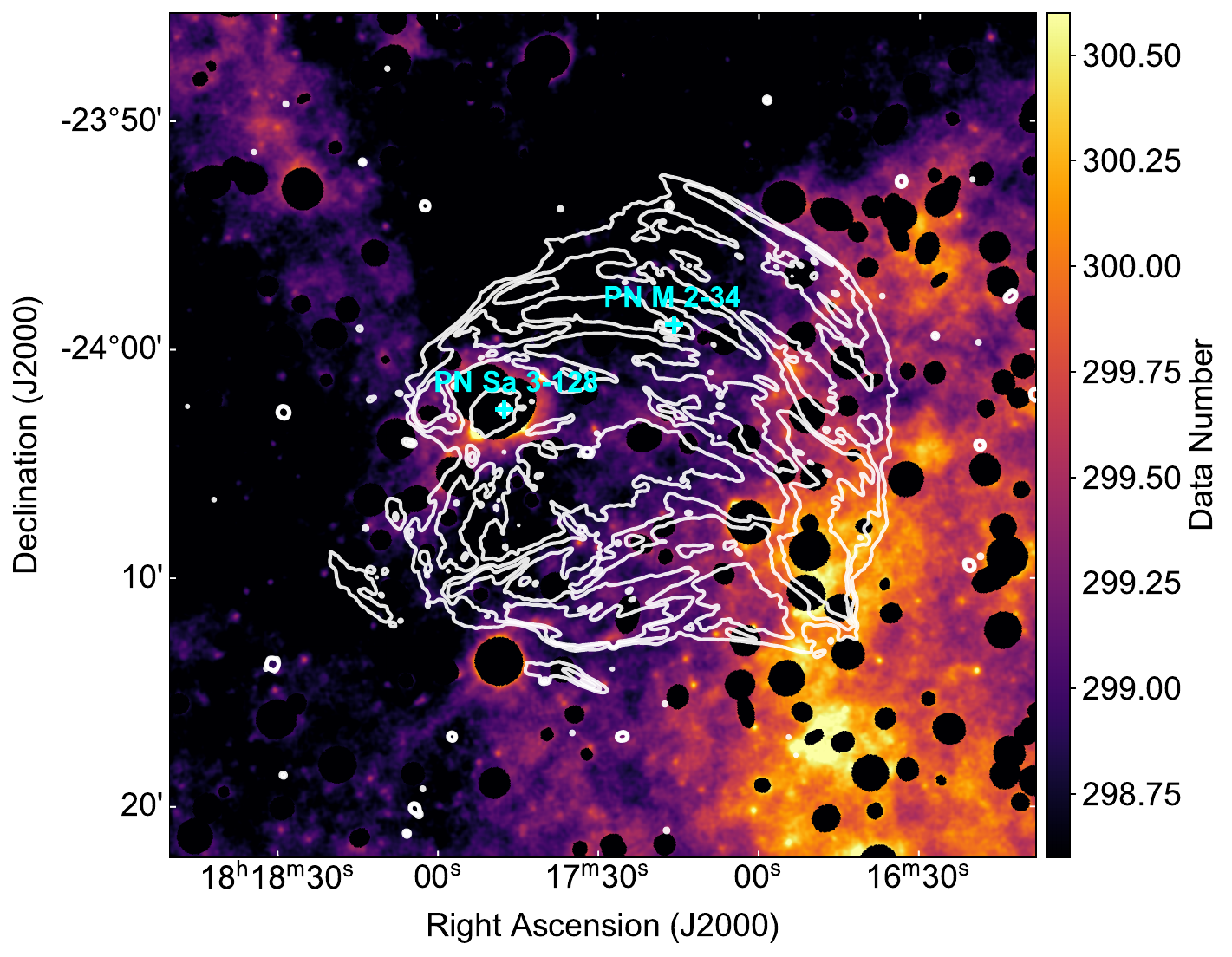}}
\caption{The WISE 22~${\rm \mu m}$ image toward G7.7$-$3.7 overlaid with white contours of 1.4 GHz total intensity levels at 0.2 mJy/beam, 0.4 mJy/beam, and 0.6 mJy/beam. The bright point-like sources are subtracted. The cyan crosses denote the two planetary nebulae named PN M 2-34 and PN Sa 3-128.
\label{fig:wise}}
\end{figure}

G7.7$-3.7$ shows a multi-shell and aspherical morphology in the radio band, with an enhancement of the radio emission in the southern shell, southwestern clump and eastern clump. 
The two radio-bright clumps are nearly unpolarized (Figure \ref{fig:PIPD}) but show steep spectra (Figure \ref{fig:spx}), indicating that the emission is nonthermal but the magnetic fields are disordered. Such properties may occur when the SNR interacts with an inhomogeneous medium, where the increased gas density lead to the bright radio emission but the magnetic fields are highly turbulent.
In addition, the bright region in the southwestern part of G7.7$-$3.7 exhibits a concave-shaped morphology, likely results from an impact with dense environmental gas.
Figure \ref{fig:wise} shows the WISE 22~${\rm \mu m}$ image, with bright point-like sources subtracted. A bright diffuse infrared emission is observed in the southwest region, coincident with the bright radio structure. This supports the existence of denser gas west of the SNR, although a line-of-sight projection effect cannot be entirely excluded.
Collectively, these properties indicate that the southwestern part of the SNR is likely interacting with a dense ambient cloud.  


\begin{figure}[ht!]
\centering
\resizebox{0.4\textwidth}{!}
{\plotone{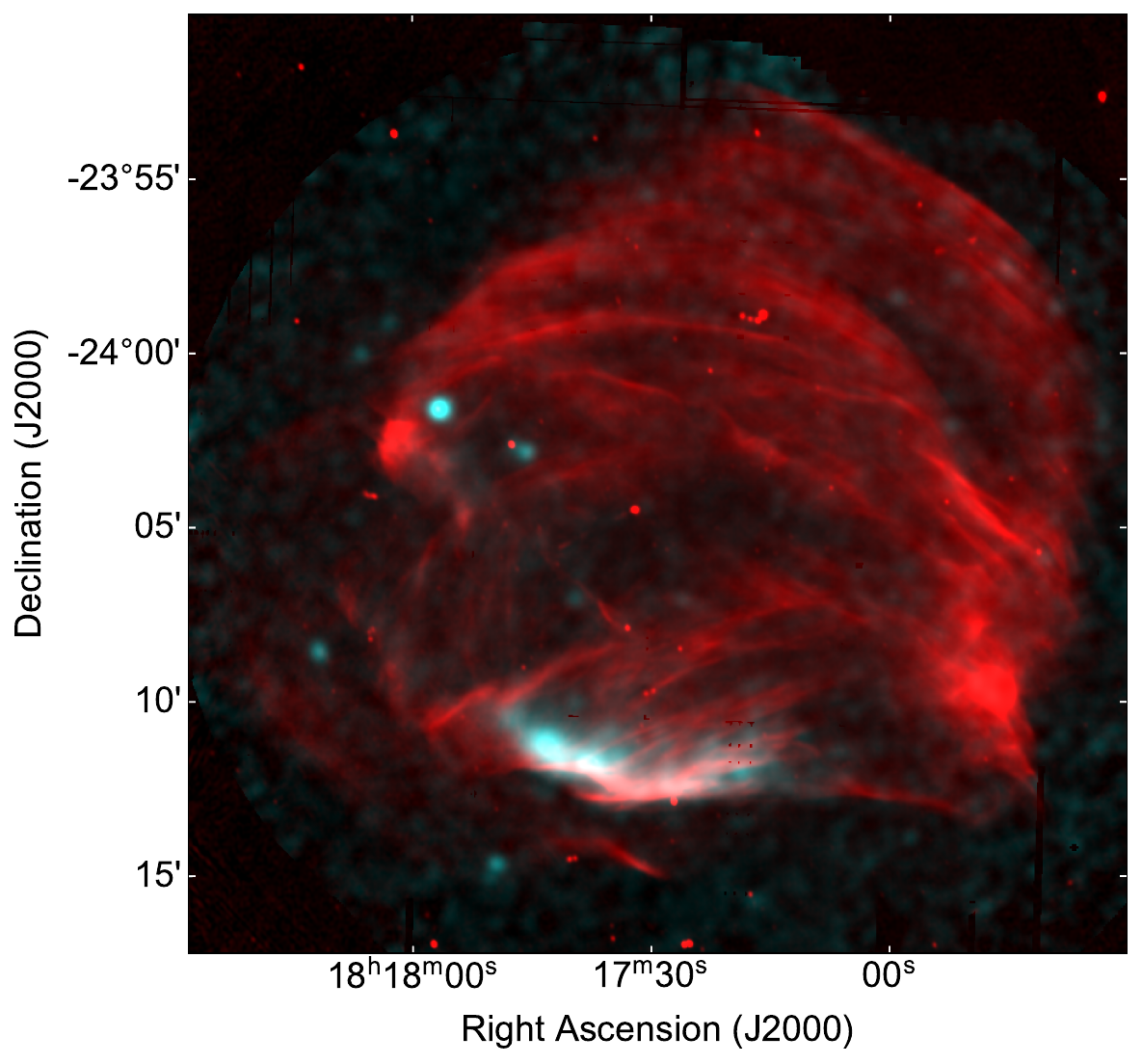}}
\caption{Composite image of G7.7$-$3.7 in VLA 1.4 GHz (red) and X-ray (cyan). The X-ray image was made using the \textit{XMM-Newton} data in 2012 (OBSID: 0671170101, \citealt{Zhou2018b}). 
\label{fig:xray+vla}}
\end{figure}

Figure \ref{fig:xray+vla} shows the composite image of G7.7$-$3.7 in VLA 1.4 GHz (red) and X-ray (cyan). Unlike radio image, the X-ray image of G7.7$-$3.7 only reveals a bright arc in the south and some very dim emission in the east. Along the south arc, the X-ray emission is bright where the radio emission is relatively faint. Previous X-ray analysis indicated that the ambient density of G7.7$-$3.7 is likely non-uniform \citep{Zhou2018b}. In addition, previous optical studies \citep{Domcek2023b} also indicated a collision with a dense circumstellar shell that lies in the southern region of G7.7$-$3.7.

\subsection{The nature of RM variation}

The RM map of G7.7$-$3.7 reveals a notable variation, which may originate from the foreground ISM or/and from the magnetized materials in the SNR. Pixels exhibiting multiple peaks in the Faraday depth spectra constitute only a small fraction of the entire region. Therefore, they do not contribute to this variation in the RM map (see details in Appendix \ref{sec:appendix}). We compared the RM map (see Figure \ref{fig:RMPA}) and the WISE 22 $\mu$m image (Figure \ref{fig:wise}) and did not find a good correlation between them. For instance, the RM value changes abruptly from 200 rad~m$^{-2}$ to 0  rad~m$^{-2}$  in the northeast, where no strong IR emission is detected, indicating that the RM variation may not result from the foreground ISM. 

Since radio pulsars are generally considered to have no intrinsic RM, their measured RMs are attributed to the foreground ISM, making them useful tools for estimating the RM contribution toward G7.7$-$3.7 \citep{Manchester1977}. 
We searched the ATNF Pulsar Catalogue \footnote{\url{https://www.atnf.csiro.au/research/pulsar/psrcat}} \citep{Manchester2005b} for pulsars located within $3^\circ$ of G7.7$-$3.7 and at distances between 3 and 6 kpc, overlapping the estimated distance to G7.7$-$3.7. The resulting pulsar sample is listed in Table \ref{table2}. At Galactic latitudes comparable to or lower than that of G7.7$-$3.7, the pulsar RMs generally decrease with increasing distance, spanning a range of $\sim 80$–$140~{\rm rad~m^{-2}}$. This range is comparable to the average RM value of G7.7$-$3.7 ($\sim 130~{\rm rad~m^{-2}}$). 
As the Galactic latitude approaches the Galactic plane, the RM becomes negative, as seen for J1811$-$2439, while G7.7$-3.7$ shows highest RM value toward this direction. The large-scale RM spatial distribution of these pulsars does not show a clear correspondence with the RM spatial distribution of G7.7$-$3.7. 
Moreover, the RM variation across the SNR is significantly larger than the pulsars at similar distances range and nearby sky area. 
The above evidence motivates us to consider the contribution of SNR materials to the RM. 
Previous radio polarization measurements have found a few SNRs with a large RM gradient due to the magnetized progenitor winds shocked by the SNRs, such as G296.5+10.0 \citep{Harvey-Smith2010} and G1.9+0.3 \citep{Luken2020}. We consider similar scenario may also be applied to G7.7$-3.7$.



\begin{deluxetable*}{cccccc}
\tablewidth{0pt}
\tablecaption{Information of pulsars near G7.7$-$3.7 \label{table2}}
\tablehead{
\colhead{Name} & 
\colhead{\begin{tabular}{c} $l$ \\ (degree) \end{tabular}} &
\colhead{\begin{tabular}{c} $b$ \\ (degree) \end{tabular}} &
\colhead{\begin{tabular}{c} Distance \\ (kpc) \end{tabular}} & 
\colhead{\begin{tabular}{c} RM \\ (${\rm rad~m^{-2}}$) \end{tabular}} & 
\colhead{Reference} 
}
\startdata
J1813$-$2621 & $5.3$ & $-4.1$ & 3.16 \tablenotemark{{\rm \dag}} & $142.2\pm3$ &  \citet{sbo+22} \\
B1819$-$22 & $9.3$ & $-4.4$ & 3.26 \tablenotemark{{\rm \dag}} & $124\pm3$ &  \citet{njkk08}\\
B1813$-$26 & $5.2$ & $-4.9$ & 3.59 \tablenotemark{{\rm \dag}} & $90\pm5$ &  \citet{hl87}\\
J1811$-$2439 & $6.6$ & $-3.0$ & 3.76 \tablenotemark{{\rm \dag}} & $-82\pm2$ &  \citet{jkk+20}\\
B1821$-$24A & $7.8$ & $-5.6$ & $5.37\pm 0.01$ \tablenotemark{{\rm \ddag}} & $82.2\pm2$ &  \citet{dhm+15}\\
\enddata

\tablenotetext{{\rm \dag}}{The distances are estimated from the dispersion measure (DM) using the YMW16 Galactic electron density model \citep{ymw17}.}
\tablenotetext{{\rm \ddag}}{The distances are estimated based on an association with Galactic globular cluster NGC 6626 \citep{Baumgardt2021}.}

\end{deluxetable*}


G7.7$-3.7$ was proposed to originate from a low-luminous core-collapse SN \citep{Zhou2018b}, whose progenitor star is probably a massive star and launched pre-supernova winds. The progenitor winds can shape not only the density environment of the SNR, but also the magnetic fields structures.
The rotation of the star can produce a tangential magnetic field component at a large distance $r$ from the star (e.g., \citealt{Lamers1999, Chevalier1994}): 
\begin{equation}
B_{T}(r)=B_{\star}(\frac{V_{\rm rot}}{V_{\infty}})(\frac{R_{\star}}{r})
\end{equation}
where $B_{\star}$ is the magnetic field on the surface of the star, $V_{\rm rot}$ is the equatorial rotational velocity, $V_{\infty}$ is the radial wind velocity, and $R_{\star}$ is the stellar radius. This equation shows that the tangential magnetic fields of stellar winds can dominate over radial magnetic fields at a large distance. The tangential magnetic field strength increases with the stellar magnetic fields, rotation, and decreases with the wind velocity. 

\citet{Harvey-Smith2010} proposed a model in which the tangential magnetic field of a red supergiant could generate an RM value of $\sim 40~{\rm rad~m^{-2}}$. 
Under the assumption of minimum energy requirement, we can estimate the magnetic field strength of G7.7$-$3.7. Adopting the distance to G7.7$-$3.7 be 4 kpc (\citealt{Pavlovic2013}; H. C. Lin et al. in prep.), the ratio of total cosmic-ray energy to the energy in relativistic electrons as $\xi\approx100$ \citep{Burbidge1959, Katz2008}, a spectral index $\alpha=-0.38$ from this work, and typical low- and high-frequency cutoffs of $10^7$ Hz and $10^{11}$ Hz (X-ray radiation of the remnant is dominated by thermal emission, \citealt{Zhou2018b}), we estimate a magnetic field of $B_{\rm eq}= 63~{\rm \mu G}$. This result is comparable to some young core-collapse SNRs \citep{Vink2020}.
Using this magnetic field strength, the remnant radius $R=11'$, the remnant radius to shell thickness ratio $x=R/\Delta R=8$ \citep{Harvey-Smith2010}, and representative values for a red supergiant and its wind, the mass-loss rate $\dot{M}=1\times 10^{-5}~M_{\odot}~{\rm yr^{-1}}$ \citep{vanLoon2010}, and $V_{\infty}=30~{\rm km~s^{-1}}$ \citep{vanLoon2010}, we roughly estimate an RM value on the shell using the \citet{Harvey-Smith2010} model:


\begin{equation}
\begin{aligned}
{\rm RM} =&30~(\frac{B_{\rm eq}}{63~{\rm \mu G}})(\frac{R}{11'})^{-1}(\frac{d}{4~{\rm kpc}})^{-1}(\frac{x}{8})^{\frac{1}{2}} \\
& (\frac{\dot{M}}{1\times 10^{-5}~M_{\odot}~{\rm yr^{-1}}})(\frac{V_{\infty}}{30~{\rm km~s^{-1}}})^{-1}~{\rm rad~m^{-2}}.
\end{aligned}
\end{equation}

\begin{figure}[ht!]
\centering
\resizebox{0.45\textwidth}{!}
{\plotone{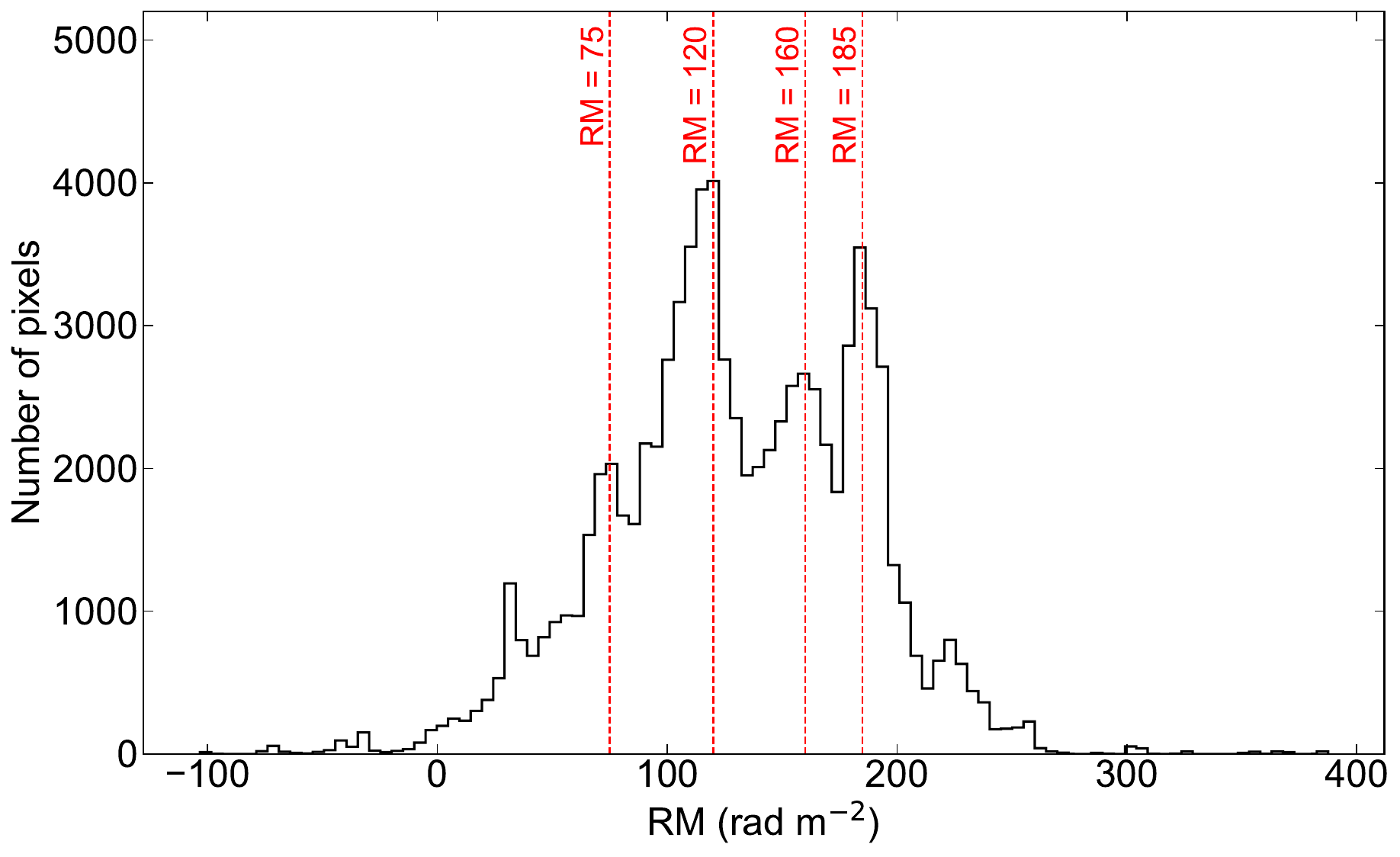}}
\caption{The RM histogram of G7.7$-$3.7. The red dashed lines represent the RM value of four significant peaks. 
\label{fig:RMhist}}
\end{figure}

The observed mean RM value of G7.7$-$3.7 is $130~{\rm rad~m^{-2}}$, similar to the foreground RM value of $80$--$140$~${\rm rad~m^{-2}}$ estimated using nearby pulsars,  with a standard deviation of $54~{\rm rad~m^{-2}}$. The RM distribution deviates from a Gaussian distribution, but shows peaks at $75~{\rm rad~m^{-2}}$, $120~{\rm rad~m^{-2}}$, $160~{\rm rad~m^{-2}}$ and $185~{\rm rad~m^{-2}}$ see Figure \ref{fig:RMhist}.

The variation of the RM values compared to the mean value is generally a few tens of ${\rm rad~m^{-2}}$, broadly agrees with the red supergiant stellar wind model of $\sim 30~{\rm rad~m^{-2}}$. However, it should be noted that this model involves many simplifications and assumptions regarding stellar properties and wind parameters, which have been unknown for G7.7$-3.7$. In addition, the reliability of the minimum energy requirement is still debated \citep{Beck2005, Urosevic2018}. More detailed investigations through future magnetohydrodynamic simulations will be essential to better quantify the RM contributions from stellar winds.

Now we consider the pattern of RM variation. 
The residual RM map after subtracting a foreground RM value of $130~{\rm rad~m^{-2}}$ reveal a pattern resembling a quadrupole configuration (see Figure~\ref{fig:schematic}). Two distinct boundaries mark the reversal in the RM sign: one in the northeast and one in the west. The northeastern boundary exhibits a rapid RM sign reversal, while the western boundary shows a smoother transition.

The simplest model of magnetized stellar wind called split monopole \citep{Sakurai1985a} can produce such a quadrupole pattern in RMs \citep{Harvey-Smith2010}. In this wind model, the magnetic polarity is opposite in the northern and southern hemispheres, separated by a current sheet (see Figure 4 a) in \citealt{Maiewski2022}). As the progenitor rotates, the pre-supernova magnetic field direction along the line of sight reverses across both the current sheet and the spin axis, resulting in an intrinsic quadrupolar reversal in the sign of the RM. 
Across the current sheet, the line-of-sight magnetic field direction reverses sharply and abruptly, whereas across the spin axis, the change is gradual and continuous. This leads us to infer that the spin axis of the progenitor may align with the western boundary of G7.7$-$3.7. 

A schematic illustration of this scenario is shown in Figure \ref{fig:schematic}. This is an oversimplified model that includes the assumption that the spin axis and magnetic axis of the progenitor star are the same. In reality, factors such as the misalignment between the magnetic and rotation axes, the viewing angle of the observers, and the density distribution of the surrounding environment can all cause the RM distributions to not strictly conform to a quadrupole pattern, as observed in G7.7$-$3.7.

The sky-projected magnetic field directions are mostly aligned with the filaments of G7.7$-$3.7 (see the right panel of Figure \ref{fig:RMPA}). Such an orientation is naturally expected from shock compression, where the component of the ambient magnetic field tangential to the shock front is amplified due to compression, leading to an ordered magnetic field preferentially oriented along the SNR shell.

\subsection{Cocoon-like morphology}
\begin{figure}[ht!]
\centering
\resizebox{0.45\textwidth}{!}
{\plotone{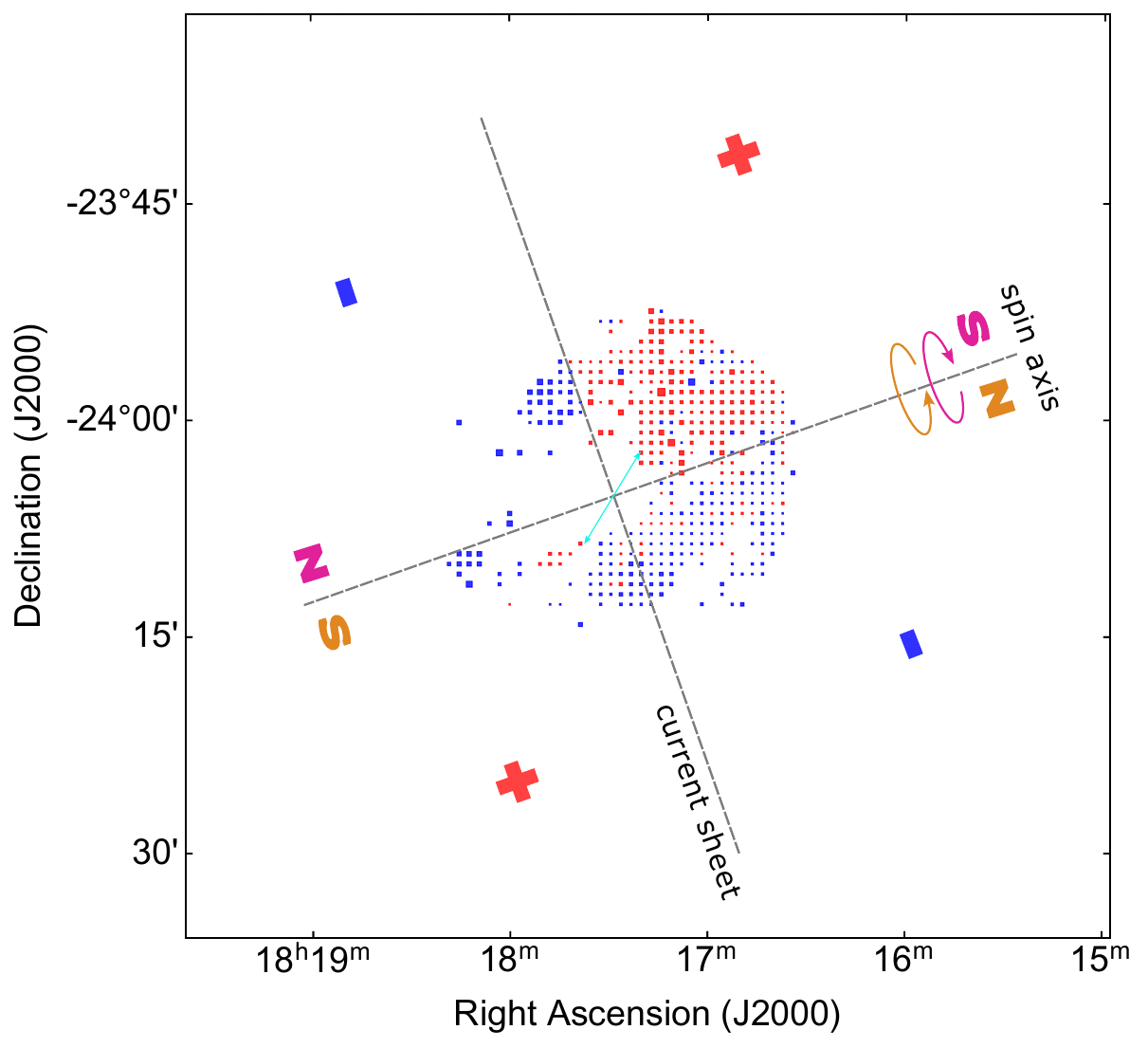}}
\caption{The RM square image of G7.7$-$3.7 overlaid with schematic illustration of the split monopole stellar wind scenario. The size of each square indicates the absolute value of the RM after subtracting an assumed foreground RM of $130~\mathrm{rad~m^{-2}}$. Red squares represent positive residual RM values, while blue squares represent negative ones. Brown and pink annotations denote the two possible magnetic polarities and the corresponding rotation directions of the progenitor star. A possible direction of motion for the progenitor star is indicated by the cyan line.
\label{fig:schematic}}
\end{figure}

Finally, we offer brief remarks on the peculiar cocoon-like morphology of G7.7$-3.7$.
We propose that the multi-shell, cocoon-like morphology is likely due to an interaction between the SNR and pre-existing shells created by its progenitor star. 
This is a more natural explanation compared to a chance alignment with a highly structured and layered ISM. 
Such multiple shells have been found in some evolved massive stars with unstable mass-loss or wind-wind collisions \citep{VanMarle2012, Decin2012, Rubio2020, Lau2022}.
The shells of the remnant do not appear to share a common center but instead are approximately coaxially aligned along a northwest–southeast axis. A possible explanation is that its progenitor star may have a motion along the northwest-southeast direction, as indicated by the cyan line in Figure \ref{fig:schematic}.
Moreover, the southern shell and the southwestern and eastern radio clumps are likely interacting with denser materials, resulting in enhanced radio emission and deformed morphology \citep{Jun1999, Sofue2024}.
While the northern shells seem to be coaxially aligned, the southern structures do not show such ordered arrangement. 
This scenario resembles the case of VRO 42.05.01 \citep{Arias2019b}, where the progenitor star is thought to have traversed regions of different densities. Future observations and simulations will be necessary to test the validity of this interpretation. 
The above interpretation thus favors a massive progenitor star for G7.7$-3.7$.

Our results cannot confirm or rule out whether G7.7$-$3.7 is a historical SNR. The integrated spectral index of G7.7$-$3.7 ($-0.38\pm 0.04$) is flatter than other historical SNRs, such as Kepler's SNR ($-0.66\pm 0.01$, \citealt{Castelletti2021a}), Tycho's SNR ($-0.62 \pm 0.01$, \citealt{Castelletti2021a}), SN 1006 ($-0.6$, \citealt{Green2025}), and RCW 86 ($-0.6$, \citealt{Green2025}). However, historical SNRs exhibit considerable diversity in their properties. For instance, RX J1713.7$-$3946 only has faint radio emission with unknown spectral index \citep{Ellison2001}, and Pa 30 even does not have radio emission detected \citep{Shao2025a}. Therefore, this result does not exclude it from being a historical SNR. Further multi-wavelength observations will be essential to clarify whether it is a historical SNR.



\subsection{Blowouts}
\begin{figure}[ht!]
\centering
\resizebox{0.45\textwidth}{!}
{\plotone{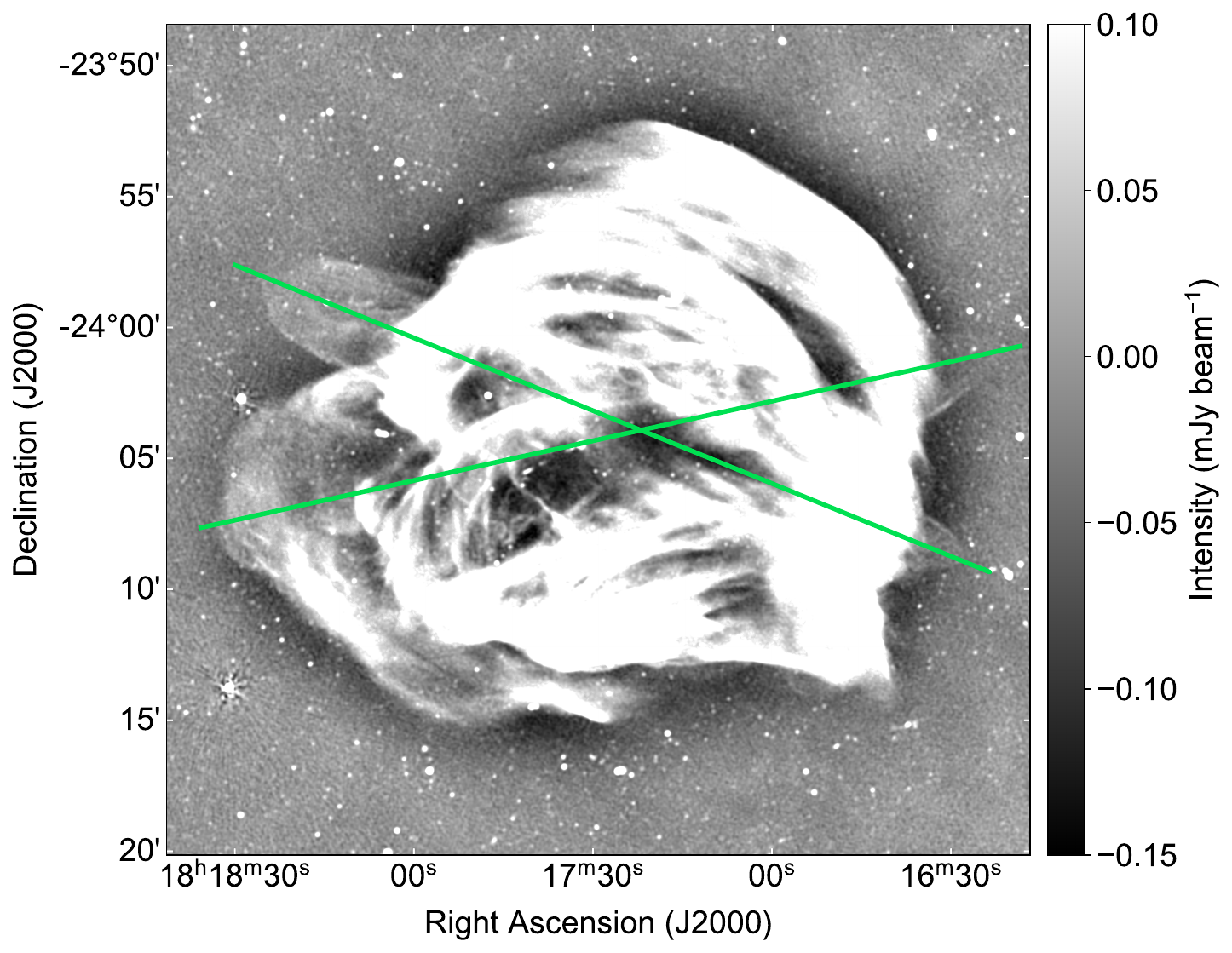}}
\caption{MeerKAT image of G7.7$-3.7$ at 1335 MHz \citep{Cotton2024}. The green solid lines connects the two opposite blowouts.
\label{fig:meerkat}}
\end{figure}

Another notable feature of G7.7$-$3.7 is its two faint blowouts in the east (See Figure \ref{fig:I}).
The smaller northeastern blowout has a bowshock-like morphology with a protrusive length of $\sim 5'$ away from the main shell.
The southeastern blowout is more extended, with a protrusive length of $\sim 6'$. 
In addition, two fainter blowouts were detected in the opposite direction of these two blowouts in the MeerKAT image (\citealt{Cotton2024}, see Figure \ref{fig:meerkat}), features that are barely visible in our VLA image. 
Although the two western blowouts exhibit a symmetrical alignment with those in the east, the morphology of them is quite different. The western blowouts are smaller in scale and sharper than the eastern blowouts.

The origin of the blowout structures may be interpreted through two possible scenarios. One possibility is that they are the result of the interaction between the SNR shock and the environment shaped by its progenitor activities.
Progenitor activities such as the nonuniform density distribution of CSM\citep{Blondin1996, Chiotellis2021a} can form two opposite blowouts. 
Alternatively, protrusions may arise from intrinsic asymmetries of the SN explosion itself, such as the anisotropic ejecta distribution or high-velocity clumped ejecta shrapnel \citep{Aschenbach1995a, Wang2002, Miceli2008, Orlando2016}. 
In addition, jets launched by the progenitor system or during the SN explosion can also form two opposite blowouts \citep{Fesen2006}.

In the case of G7.7$-$3.7, the northeastern blowout shows a bow-like morphology, which could be the result of a jet-like outflow or ejecta shrapnel. The orientation of this pair of blowouts is not aligned with the spin axis of our above scenario (see Figure \ref{fig:schematic}). While jets produced by the progenitor system may align with the spin axis of the progenitor, jets may also be generated during the supernova explosion itself \citep{Fesen2006}. In such cases, the jet direction is not necessarily required to align with the spin axis of the progenitor. Ejecta shrapnel is often expected to exhibit X-ray emission due to shock-heated metal-rich material \citep{Aschenbach1995a, Miceli2008, Wang2002}, yet no corresponding X-ray structures have been reported for these blowout regions \citep{Zhou2018b}. In addition, ejecta shrapnel does not necessarily produce paired oppositely directed blowout structures, which seems to be the case for G7.7$-$3.7. 

In contrast, The southeastern blowout appears more diffuse, making it less likely to be the result of a jet or ejecta shrapnel. The influence of the CSM density distribution may be a more possible origin of this blowout. 



\section{Conclusion} \label{sec:conclusion}
We performed new L-band observations of SNR G7.7$-$3.7 using VLA in its C and B-configurations, providing a high-resolution radio view of this remnant. The 1.4 GHz continuum map reveals a cocoon-like morphology consisting of multiple shells and faint blowouts. We measured a flux density of $9.6\pm 0.5$ Jy at 1.4 GHz for G7.7$-$3.7. The spectral index map of G7.7$-$3.7 shows predominantly nonthermal emission, with an integrated spectral index of $-0.38\pm 0.04$. 
Under the minimum-energy assumption, we estimate a magnetic field strength of $\sim 63~{\rm \mu G}$.
This value is typical in core-collapse SNRs. 

Polarization images of G7.7$-$3.7 show high linear polarization fraction ($30\%$--$40\%$) in the northwestern filaments and moderate polarization ($10\%$--$20\%$) in the northeast and south. The sky-projected magnetic field directions are mostly aligned with the filaments of G7.7$-$3.7. These results indicate that the magnetic fields are relatively ordered in these filaments and the magnetic field orientation agrees with a compression origin. 

Radio-bright clumps in the southwestern and eastern parts of G7.7$-3.7$ are nearly unpolarized but show steep spectra. In particular, the southwestern bright region has a concave morphology that coincides with infrared enhancements, suggesting the existence of a denser gas west of G7.7$-3.7$, although a line-of-sight projection effect cannot be completely excluded.


The RM distribution exhibits a large variation across the remnant. 
We infer that the multi-shell radio morphology and the large RM variation across G7.7$-3.7$ can be explained by an interaction with magnetized progenitor winds. In addition, the progenitor may have a motion along the northwest to the southeast through two media of different densities. This result suggests a core-collapse SN origin of this SNR.

\begin{acknowledgments}

We are grateful to Lingrui Lin for his assistance in data reduction and extend our thanks to Wolfgang Wreich, Patricia Reich and Yi-Heng Chi for their valuable comments.
We thank the anonymous reviewer for the useful comments that improved this paper.
T.-X.L. and P.Z. acknowledge the support from the National SKA Program of China (No.2025SKA0140100), National Natural Science Foundation of China (NSFC) grant No.\ 12273010 and the Fundamental Research Funds for the Central
Universities with grant No. KG202502.
C.-Y. N. and S. Zhang are supported by a GRF grant of the Hong Kong Government under HKU 17304524.
We used the Large Language Model DeepSeek-V3.2 for language refinement in the manuscript. All scientific details were written and verified by the authors.
\end{acknowledgments}





%
\facilities{VLA}

\software{CASA \citep{Team2022}, RM-Tools \citep{Purcell2020}, Aegean \citep{Hancock2012, Hancock2018a}, astropy \citep{2013A&A...558A..33A,2018AJ....156..123A,2022ApJ...935..167A}, CARTA \citep{Comrie2021}, SAOImage DS9 \citep{Joye2003}, Starlink \citep{Currie2014, Berry2022}, SciPy \citep{2020SciPy-NMeth}}


\appendix
\counterwithin{figure}{section}

\section{Faraday depth spectra} \label{sec:appendix}

We extracted the Faraday depth spectrum from each pixel to check if there are multiple components along the line of sight. 
The Peaks in the spectra were identified using the \texttt{find\_peaks} function from the SciPy package.
The detection threshold was set to $5\sigma_{\rm MAD}$, where $\sigma_{\rm MAD}$ is the estimated noise in the Faraday depth spectrum, calculated using the median absolute deviation of the polarized intensity in the spectrum following the algorithm described in \citep{Purcell2020}\footnote{\url{https://github.com/CIRADA-Tools/RM-Tools/wiki/RMsynth1D}}. The separation between adjacent peaks was required to be larger than $64~{\rm rad~m^{-2}}$, corresponding to the FWHM of the RM spread function (RMSF). In addition, the polarized intensity of the second-highest peak was required to exceed $50\%$ that of the highest peak. Otherwise, this spectrum was identified as having only one peak. 

\begin{figure*}[ht!]
\centering
\resizebox{0.98\textwidth}{!}
{\plotone{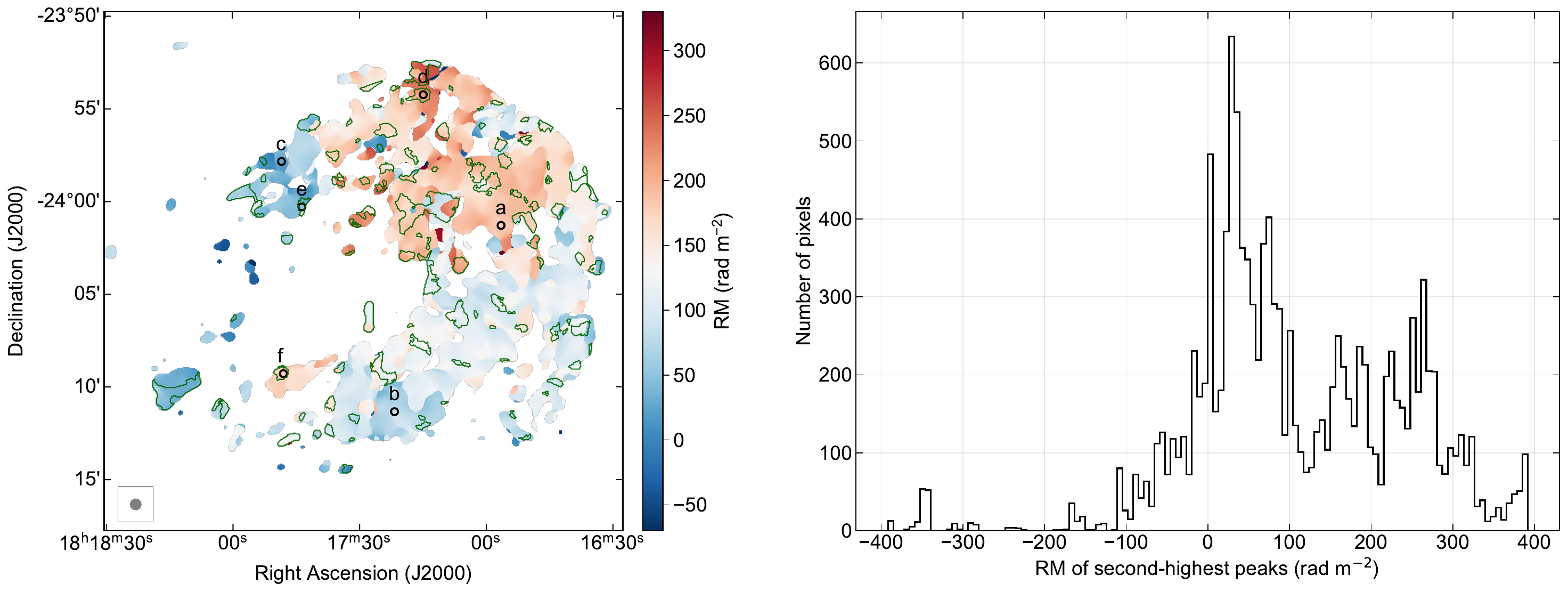}}
\caption{\textit{Left:} The RM map of G7.7$-$3.7 overlaid with green regions where multiple peaks are detected in the Faraday depth spectra. For the pixels labeled with a--f, the Faraday depth spectra were extracted from the cube and are shown in Figure \ref{fig:FDF}. \textit{Right:} Histogram of the RM values corresponding to the second-highest peaks in the Faraday spectra.
\label{fig:doublepeak}}
\end{figure*}

\begin{figure*}[ht!]
\centering
\resizebox{0.95\textwidth}{!}
{\plotone{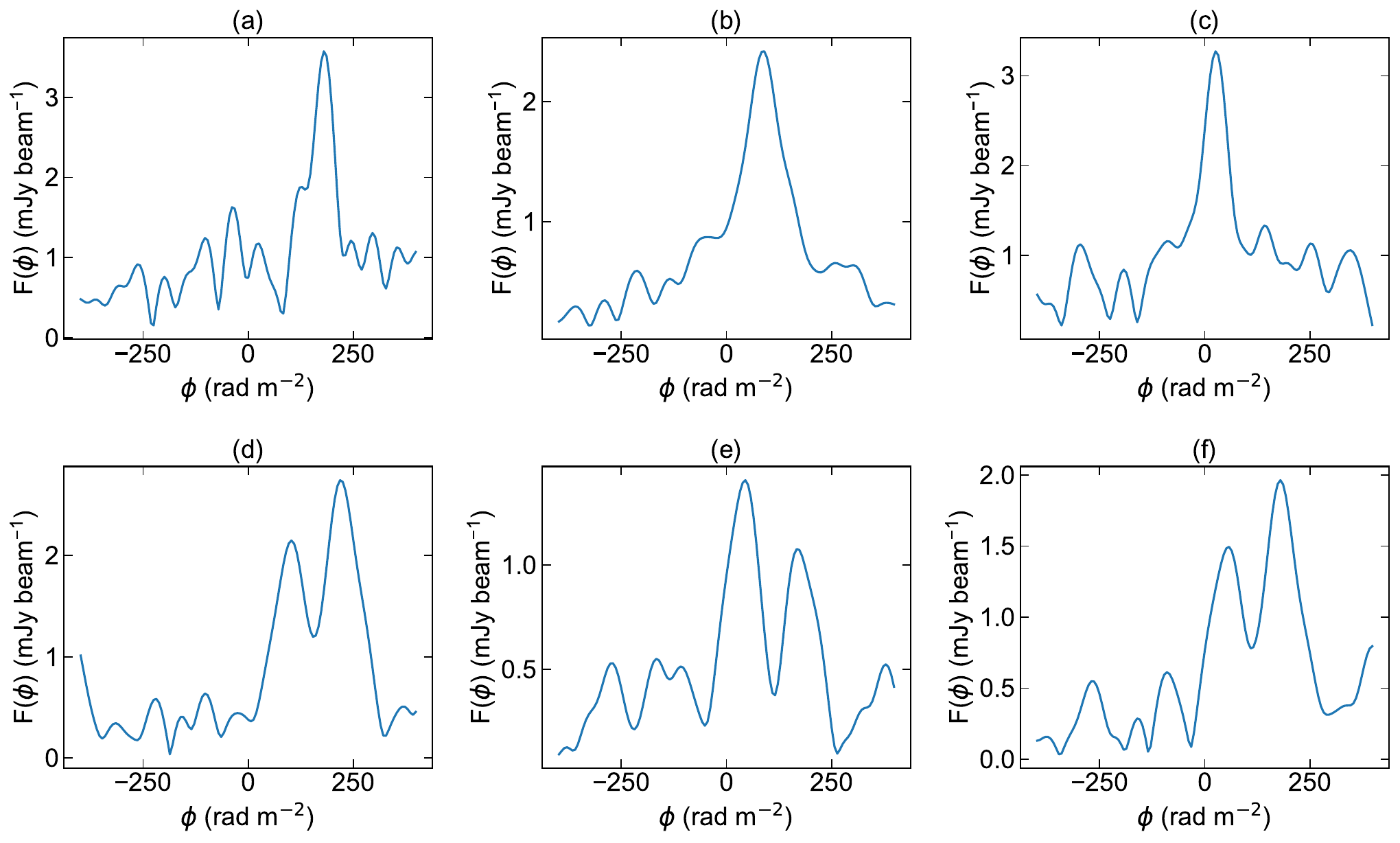}}
\caption{Faraday depth spectra for pixels a--f in the left panel of Figure \ref{fig:doublepeak}.
\label{fig:FDF}}
\end{figure*}

Using the above criteria, we identified regions with multiple peaks in the Faraday depth spectra (see the left panel of Figure \ref{fig:doublepeak}). 
Examples of spectra with single and multiple components are presented in Figure \ref{fig:FDF}.
The spatial distribution of pixels showing multiple peaks does not reveal any obvious pattern. 
Moreover, such pixels account for only a small fraction of the data compared to the four dominant components identified in the RM histogram (Figure \ref{fig:RMhist}; see also the right panel of Figure \ref{fig:doublepeak}).
We therefore suggest that the multiple peaks may come from local turbulence of the SNR. Distinct turbulent units along the line of sight may produce separate peaks in the Faraday spectrum. 
In addition, some secondary peaks may be related to sidelobes of the RMSF, since the Faraday depth spectra have not been deconvolved. In either case, these features do not contribute to the large-scale variation pattern in the RM map.

\bibliography{sample7}{}
\bibliographystyle{aasjournalv7}



\end{document}